# Laser Liftoff of GaAs Thin Films

Garrett J. Hayes*, and Bruce M. Clemens
Department of Materials Science and Engineering, Stanford University, Stanford CA, 94305
*email:  ghayes@alumni.stanford.edu

**Abstract:** The high cost of single crystal III-V substrates limits the use of GaAs and related sphalerite III-V materials in many applications, especially photovoltaics.  Separating epitaxially-grown layers from a growth substrate can reduce costs, however the current approach, which uses an acid to laterally etch an epitaxial sacrificial layer, is slow and can damage other device layers.  Here, we demonstrate a new approach that is orders of magnitude faster, and that enables more freedom in the selection of other device layers.  We show damage-free removal of an epitaxial single crystal GaAs film from its GaAs growth substrate using a laser that is absorbed by a smaller-band-gap, pseudomorphic layer grown between the substrate and the GaAs film. The liftoff process transfers the GaAs film to a flexible polymer substrate, and the transferred GaAs layer is indistinguishable in structural quality from its growth substrate.

GaAs and related sphalerite III-V materials are attractive for a variety of electronic and optoelectronic devices, however the high cost of III-V substrates hinders their use in certain applications, especially photovoltaics.  Separating an epitaxially-grown film from a III-V substrate and recycling the substrate is one means to reduce costs, and today this is achieved by using an acid solution to laterally etch a sacrificial layer placed between the growth substrate and the epitaxial layer(s) of interest.  This approach, however, is hindered by long etch times and possible damage to completed devices due to long immersion in etchant solution[1-3]. Here, we introduce a new approach, based on spatially controlled energy deposition from a pulsed laser, that separates epitaxial single crystal sphalerite III-V films from their III-V growth substrates. It is orders of magnitude faster, does not require long immersion in etchant solution, and allows release of completed devices. We demonstrate this approach by using a common 1064 nm, nanosecond, Q-switched laser to transfer a single crystal epitaxial GaAs film from its GaAs growth substrate to a flexible polymer substrate. Furthermore, we show that this transferred GaAs layer is indistinguishable in structural quality from its single crystal GaAs growth substrate.

To achieve differential absorption of the 1064 nm light, we deposit a thin, pseudomorphic layer of InGaAsN[4-6], followed by the deposition of the epitaxial GaAs film – which, due to the absence of misfit dislocations in the InGaAsN layer, is of exceptionally high structural quality. We then adhere a flexible polymer substrate to the GaAs film surface before exposing the structure to a single laser pulse directed through the GaAs substrate. By tuning the composition of the InGaAsN layer such that its bandgap is lower than 1.165 eV (energy of a 1064 nm photon), the InGaAsN layer strongly absorbs 1064 nm laser light to which the GaAs substrate is effectively transparent. Upon absorption of the laser pulse, ablation occurs along the InGaAsN layer, separating the GaAs film from its GaAs growth substrate, producing a crack-free GaAs layer adhered to a flexible polymer substrate.

Using grazing incidence X-ray diffraction (XRD), we show that the as-grown epitaxial GaAs film is indistinguishable in structural quality to its GaAs growth substrate. Then, using cross sectional transmission electron microscopy (TEM), we show that the InGaAsN and GaAs layers contain no detectable quantity of dislocations before liftoff; and, that no detectable quantity of dislocations are introduced into the GaAs film as a result of laser liftoff. Lastly, we provide preliminary data on a simple, inexpensive chemical etching procedure designed to restore the GaAs growth substrate for reuse. Our results strongly suggest that with optimization of process parameters, our laser liftoff technique can be used to rapidly transfer large area sphalerite III-V films – as well as completed devices – from their III-V growth substrates to flexible substrates; and, that their growth substrates can be reused.

**Results**

Using metallorganic chemical vapor deposition (MOCVD), a 190 nm thick InGaAsN layer was deposited onto a 500 µm thick <001> GaAs substrate, followed by the deposition of a 2.25 µm thick GaAs layer (see Methods section). Photoluminescence data of this structure, as well as an XRD reciprocal space map of the InGaAsN and GaAs (115) peaks of this sample, are shown in the supplementary section (Figure S1 and S2, respectively). Using these data in conjunction with the Band Anticrossing model parameters presented by R. Kudrawiec[7], the composition of the InGaAsN layer was approximated to be $In_{0.09}Ga_{0.91}As_{0.962}N_{0.038}$.

To determine the structural quality of the GaAs film before liftoff, a grazing incidence reciprocal space map was taken of the GaAs film (113) peak, and was compared to an identical measurement taken on a bare GaAs substrate sawn from the same ingot as the growth substrate. The X-ray scattering geometry used for these measurements is shown in Figure S3. From these reciprocal space maps (Figure S4), the rocking curves shown in Figure S5 were extracted. From Figure S5, we see that the two GaAs (113) rocking curves nearly overlap, each having a FWHM of 0.003° in ω, showing that the epitaxial GaAs film is indistinguishable in structural quality to the bare GaAs single crystal substrate.

Sections of GaAs film were transferred to 3M Scotch® Magic™ Tape (cellulose acetate film with an acrylate adhesive) using the process schematically shown in Figure 1 using a Q-switched Nd:YAG laser with a pulse duration FWHM of 8-9 ns. A fluence profile of this beam is shown in Figure S6. It is worth pointing out that large area, crack free layers of GaAs can be transferred despite our highly inhomogeneous laser beam – from Figure S6, we see that fluence varies by a factor of ~4 between the lowest fluence regions and the highest fluence regions of the beam. Liftoff is achieved over a large range of average fluences from ~0.6 J/cm$^2$ to ~3.5 J/cm$^2$, although, when average fluences greater than ~0.8 J/cm$^2$ are used, sparsely scattered specks of surface damage are observed on the back surface of the GaAs substrate where the highest fluence regions of the beam impinge on the substrate – a consequence of using a highly inhomogeneous beam. Average fluence values were calculated by measuring the total pulse energy with a pyroelectric energy detector, and dividing this value by the cross sectional area of the entire beam, which was calculated by measuring the diameter of the beam's burn paper pattern.

Figure 2 shows optical images of a 2 mm x 3.4 mm section of GaAs film transferred to Scotch® tape, and the substrate from which it was removed. For this sample, a ~6.4 mm diameter beam with an average fluence of ~0.75 J/cm$^2$ was used. Careful inspection of the GaAs film reveals that it is completely free of cracks or pinholes. The newly exposed surfaces of the GaAs film, and of the substrate, are roughened due to a thin layer of material that appears to have melted and resolidified, producing surface features typically hundreds of nanometers tall. Scanning electron micrographs of these features are shown in Figure 3. The patterns seen in the optical images of the newly exposed surfaces in Figure 2 are due to an optical effect caused by variation in the spacing and morphology of these surface features due to spatial variation in the laser beam fluence. Currently, 2 mm x 3.4 mm is about the largest area of film we can transfer using our inhomogeneous laser beam, as it is the largest area that can be irradiated with no portion of the beam below the fluence threshold at which liftoff occurs. To transfer larger areas, a higher energy laser, ideally one with greater beam homogeneity, should be employed.

Bright-field cross sectional TEM images of the as-grown structure, and of the post-liftoff GaAs film, are shown in Figure 4. Both cross sections were prepared via ion milling in a dual beam focused ion beam/scanning electron microscope (FIB/SEM), as described in the Methods section. Both TEM images were taken with a beam voltage of 200 kV at an angle slightly off a <110>-type zone axis. The broad intensity variations seen in 4(**a**) and 4(**b**) are bend contours, and the vertical streaks in 4(**b**) are curtaining artifacts from the FIB sample preparation process caused by the irregular surface features in the resolidified melt region.  In 4(**b**) we see that the melted region does not penetrate deeply into the GaAs layer, which maintains a nominal thickness of 2.25 µm.  During TEM analysis, no dislocations were observed in either the GaAs or InGaAsN layers in the as-grown structure, nor were they observed in the post-liftoff GaAs film.  Multiple regions of each TEM sample were viewed at various magnifications and degrees of sample tilt, and no dislocations were detected in either sample.  The absence of dislocations in the as-grown TEM sample is consistent with the narrow XRD rocking curve of the as-grown GaAs (113) film peak; and, the absence of dislocations in the post-liftoff GaAs sample shows that the laser liftoff process does not introduce a detectable quantity of dislocations into the GaAs layer.

In order to reuse the GaAs substrate, the GaAs wafer surface must be restored to its original clean and smooth condition after the laser liftoff process.  In an attempt to facilitate this, a structure similar to the <001>GaAs substrate/InGaAsN (190 nm)/GaAs (2.25 µm) structure was grown, but with the addition of a ~1 µm thick lattice-matched InGaP layer placed between the GaAs substrate and the InGaAsN layer.  The InGaP layer was used because phosphide materials are readily etched by hydrochloric acid (HCl), which is an acid that does not etch arsenide materials, making the InGaP layer a selective sacrificial etching layer[3, 8].  Furthermore, InGaP has a bandgap greater than 1.165 eV, making it transparent to 1064 nm photons.  After removal of the GaAs layer via laser liftoff, the GaAs substrate surface was rough, as shown in Figure 5(**a**).  The substrate was then immersed in 12.1 M HCl for approximately 5 minutes. Upon removal from HCl, the substrate surface was markedly smoother, but was visibly coated with debris. The substrate was then immersed in acetone and scrubbed with a cotton swab, which removed nearly all of this debris, resulting in the mirror-like surface shown in 5(**b**). It is likely that this debris is comprised of residual InGaAsN material, which due to its very low N content, is essentially an arsenide, and is thus expected to behave chemically as an arsenide and resist etching by HCl. The rough morphology of the post-liftoff substrate, as well as the speed at which the InGaP layer was etched, suggest that this residual InGaAsN does not prevent the HCl from reaching the underlying InGaP layer, but that debris from the insoluble InGaAsN remains on the GaAs wafer surface.

Though most of the debris was cleared by scrubbing the substrate in acetone, the presence of insoluble particles in the etchant solution could be an obstacle for preparing a surface for epitaxial growth. Ideally, the particles would be prevented from forming in the first place. To do that, a multi-part chemical etch could be employed, which is the subject of a future study. With a multi-part etch, the residual InGaAsN could be etched with an arsenide-specific etchant such as $C_6H_8O_7$:$H_2O_2$:$H_2O$[8], with the InGaP layer serving as an etch stop. Next, the InGaP layer could be etched with a phosphide-specific etchant such as HCl, with the GaAs substrate serving as an etch stop[8]. The same scheme could also be employed to the GaAs film by placing a second InGaP layer between the InGaAsN layer and the GaAs film, which would result in a high quality GaAs film surface that would allow for additional processing of the GaAs film after liftoff, if desired. Since these steps are not lateral etches, they would be fast, meaning the samples would spend very little time in the etchant solutions, which would help prevent damage to other device layers caused by the etchant solution.

The results presented herein raise new fundamental materials science questions. The atomic-scale mechanisms underlying the liftoff process are still unknown and are relatively unexplored at this time. From a fundamental perspective, there is no atomic-level understanding of the structural processes that occur when a buried single crystal layer is selectively superheated on short time-scales. A number of questions arise: How does the structure of the buried, coherently-strained InGaAsN layer evolve as it is superheated above its melting point in a confined geometry? Furthermore, without a free surface to nucleate the liquid phase, what is the nature of this phase transition and how fast does it evolve? Answering these questions will not only provide useful insight for the optimization of our laser liftoff process, but may also dramatically expand our understanding of some of the most fundamental concepts in materials science.

In summary, our laser liftoff approach provides a new means for transferring high quality single crystal sphalerite III-V films from their growth substrate to a flexible polymer substrate with no deterioration in crystalline quality of the layers during transfer. Our approach provides a new pathway for creating flexible III-V devices that overcomes the major challenges faced by the lateral etching approach. With increased options in choosing other device layers, flexible multilayer III-V devices, including flexible multijunction solar cells, may someday become a reality using this process. Using a liftoff fluence of 0.75 $J/cm^2$, a commercially available 10 J Nd:YAG nanosecond pulsed laser would provide enough energy to separate over 13 $cm^2$ of film in a single pulse – a sufficiently large area for a practical solar cell.  By syncing multiple lasers to create a single large pulse, even larger areas of film could be separated.

**Methods**

Using MOCVD, the 190 nm thick InGaAsN layer was deposited at 520°C and 100 mbar reactor pressure using an In/Ga molar flow ratio of 0.11, a N/V ratio of 0.97, and a III/V ratio of 0.0028, followed by the 2.25 µm thick GaAs layer using a III/V ratio of 0.074, also at 520°C and 100 mbar.  The precursors for In, Ga, As, and N were trimethylindium, trimethylgallium, tertiarybutylarsine, and 1,1-dimethylhydrazine, respectively.  The InGaP layer, described in the substrate reuse experiment, was deposited at 520°C and 100 mbar using an In/Ga molar flow ratio of 0.42, a III/V ratio of 0.058, using tertiarybutylphosphine as the phosphorus precursor.  For the as-grown TEM sample, the GaAs film cross-section was milled from a sample of the structure comprised of: <001>GaAs substrate/InGaAsN (190 nm)/GaAs (2.25 µm).  For the post-liftoff GaAs film TEM sample, a piece of the aforementioned heterostructure was adhered to a Si substrate along the GaAs film surface with CrystalBond™ adhesive, and was then irradiated through the GaAs substrate with a single laser pulse of fluence 1.195 J/cm$^2$, thereby transferring the GaAs film to the Si substrate.  A cross section of this transferred GaAs film was then ion-milled from the Si substrate.

## Acknowledgements

The authors would like to thank the National Science Foundation, and the ARCS Foundation, for supporting Garrett Hayes through fellowships; David Susnitzky, Fangfang Mao, and Robert Mendez of Evans Analytical Group in Sunnyvale, CA for their work in preparing and imaging the TEM cross-sections; Jonny Goodfellow and Karel Urbanek for assisting with the laser experiments; Grey Christoforo for thermal modeling; Xiaoqing Xu and Vijay Parameshwaran for MOCVD assistance; Robert Chen for his expertise and assistance with the PL measurements; Tomas Sarmiento for assistance with the Band Anticrossing model calculations; Bob Hammond for his mentorship; and Alberto Salleo for continual guidance and free roam of his laser lab.

## Author contributions

The authors conceived of the ideas together, and Garrett Hayes performed the experiments and wrote the manuscript.

## Additional information

**Competing financial interests:** The authors declare no competing financial interests.

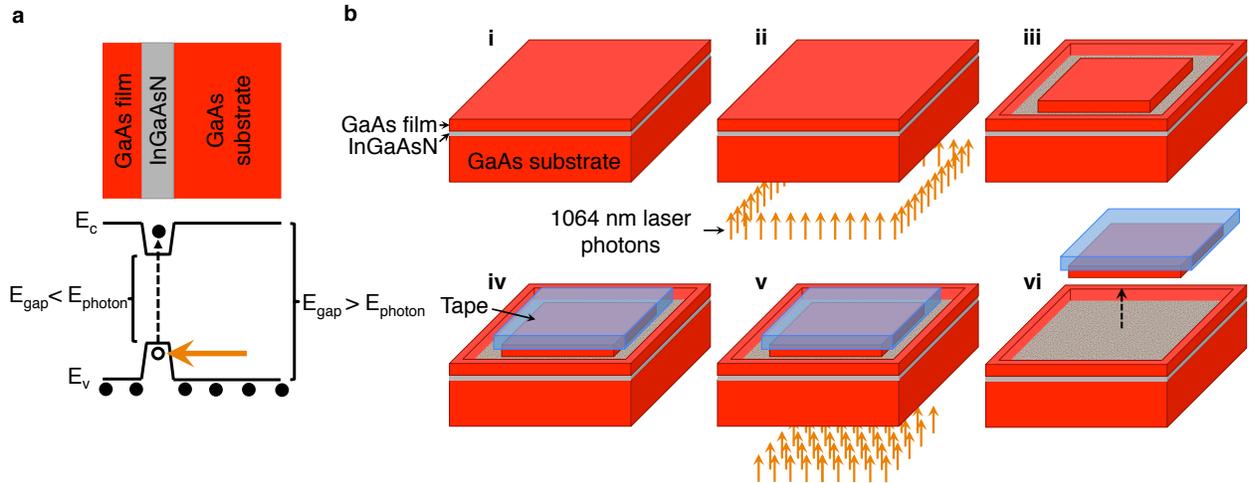

**Figure 1: GaAs laser liftoff schematic. a**, Band diagram of the GaAs/InGaAsN/GaAs structure, showing a photon passing through the GaAs substrate into the InGaAsN layer where it is absorbed. **b**, Laser liftoff process flow: **i**, A GaAs substrate/InGaAsN/GaAs structure is prepared. **ii**, A region of GaAs film is isolated from adjoining GaAs film by irradiating the perimeter with a 1064 nm laser pulse. This can be accomplished using a simple shadow mask consisting of two parallel razor blades, forming a slit-like beam. Typically, to isolate 4 edges, the sample is irradiated four times, irradiating one edge at a time, repositioning the beam each time to define an edge. Upon exposure, GaAs film is ejected from the substrate. When all four edges have been isolated, the result is the structure shown in (**iii**). **iii**, Shows a region of GaAs film with isolated edges. **iv**, Tape is adhered to the surface of the isolated GaAs film. **v**, The isolated GaAs film is irradiated with a single 1064 nm laser pulse, resulting in the ejection of the GaAs film from the GaAs substrate. **vi**, The final result – a crack free GaAs film adhered to tape.

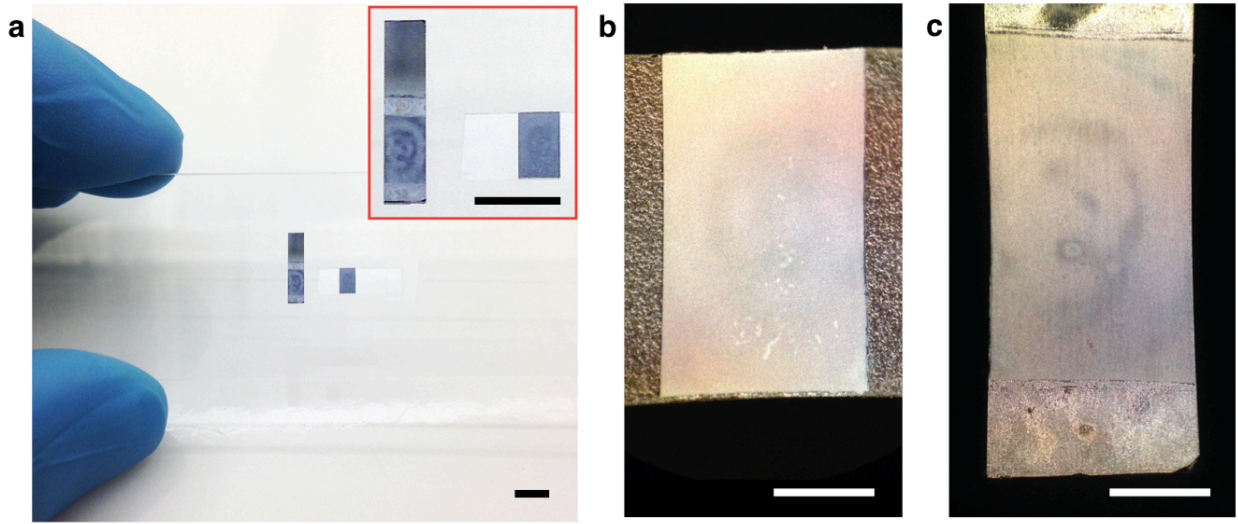

**Figure 2: Single crystal GaAs thin film on tape: a**, Image of a 2 mm x 3.4 mm section of GaAs film (2.25 µm thick) adhered to Scotch® tape (right), and the substrate from which it was removed (left). Inset shows a magnified view. The GaAs film contains no cracks or pinholes. **b**, Dark-field optical microscope image of the GaAs film after liftoff. **c**, Dark-field optical microscope image of the substrate after liftoff. Prior to liftoff, the top and bottom edges of the GaAs film were isolated using the procedure described in Figure 2, however, the left and right edges were cleaved edges, and thus did not require this procedure. Scale bars: 4 mm (**a,** inset (**a**)), 1 mm (**b**, **c**)

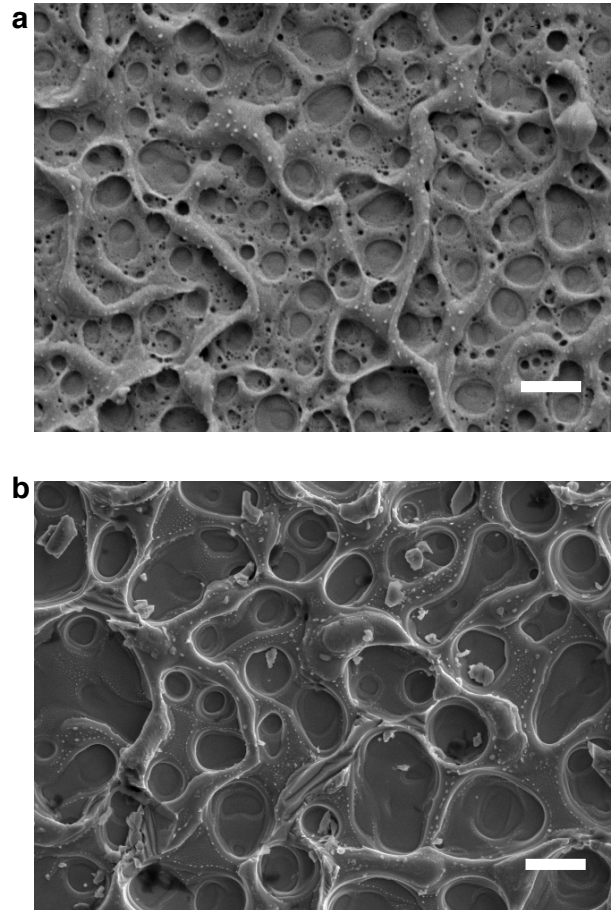

**Figure 3: Film and substrate separation surfaces. a**, Scanning electron micrograph of the separation surface of a 2.25 µm thick GaAs film that was separated from its GaAs growth substrate and transferred to Scotch® tape using an average beam fluence of ~0.75 J/cm². **b**, The substrate from which the film in (**a**) was separated. Scale bars: 2 µm (**a**, **b**)

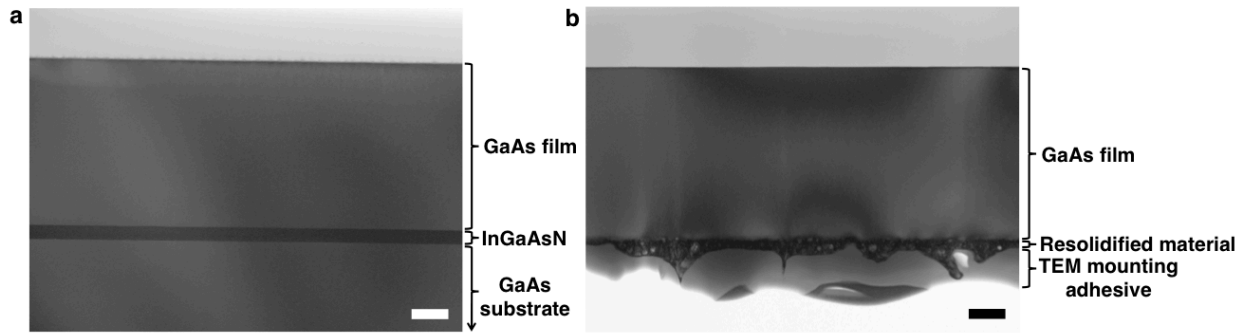

**Figure 4: GaAs film TEM cross sections – before and after liftoff**. **a**, Cross sectional bright-field TEM image of the as-grown <001>GaAs substrate/InGaAsN (190 nm)/GaAs (2.25 µm) heterostructure. **b**, Cross sectional bright-field TEM image of the post liftoff GaAs film. The vertical streaks in (**b**) are curtaining artifacts[12] from the FIB sample preparation process caused by the irregular surface features in the melted region. The broad intensity variations seen in both samples are bend contours. Both samples were prepared via ion milling using a FIB/SEM and both images were taken with an accelerating voltage of 200 kV at an angle slightly off the <110>-type zone axis. Both cross sections are free of dislocations. Scale bars: 500 nm (**a**, **b**).

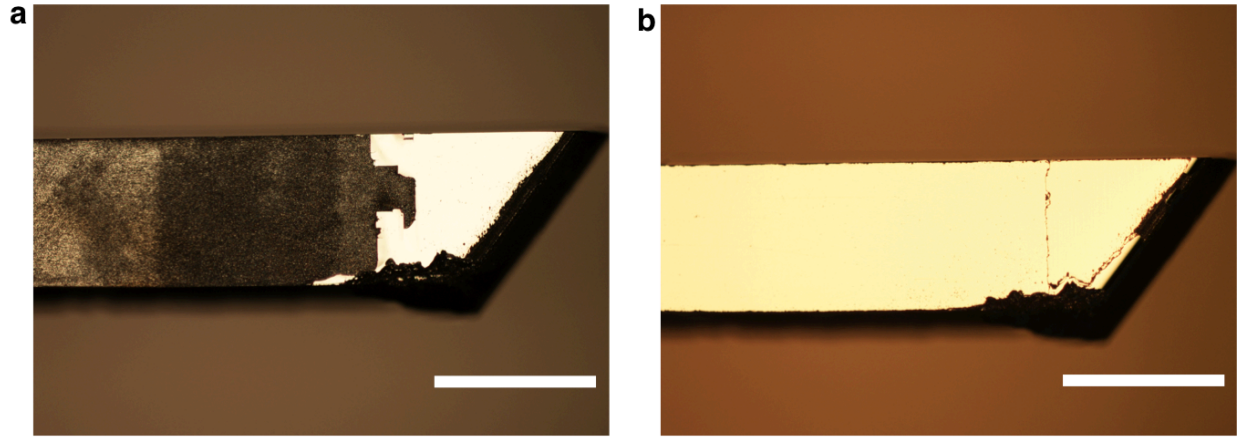

**Figure 5: Post-liftoff growth substrate, before and after HCl cleaning: a**, Bright-field optical microscope image of the InGaP-coated GaAs growth substrate after liftoff of the GaAs film. **b**, Image of the same substrate after it was placed in 12.1 M HCl for 5 minutes, then immersed in acetone and scrubbed with a cotton swab. Scale bars: 1 mm (**a**, **b**).

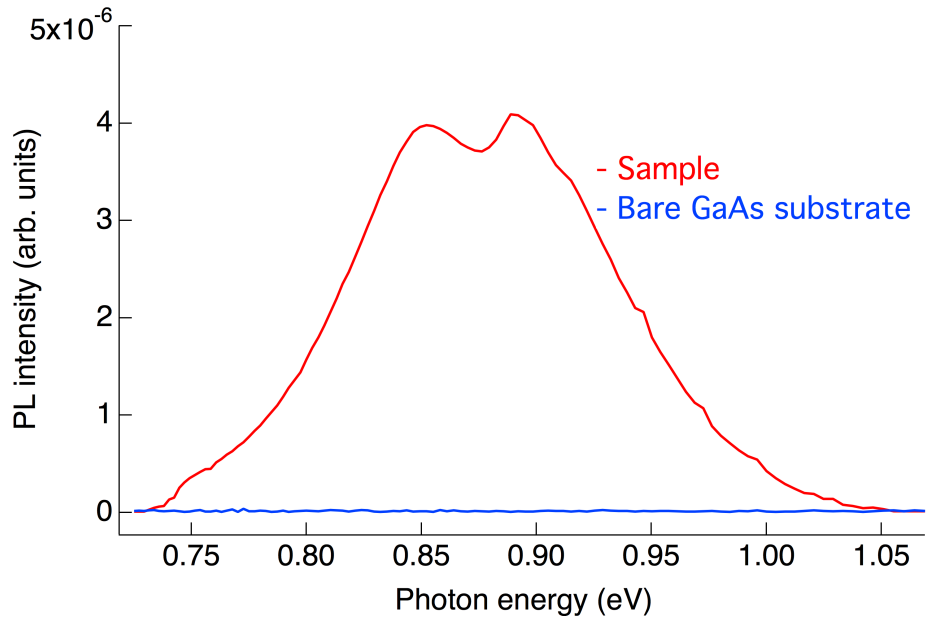

**Figure S1: Photoluminescence.** Photoluminescence intensity versus photon energy for the <001>GaAs substrate/InGaAsN (190 nm)/GaAs (2.25 µm) heterostructure, as well as for a bare GaAs substrate sawn from the same ingot as the growth substrate. The strong emission peak centered at 0.87 eV corresponds to the bandgap of the InGaAsN layer. Measurements were performed using a continuous wave 60 mW chopped laser with a wavelength of 980 nm combined with a Newport 818IG detector lock-in setup with a 1100 nm long pass filter placed before the spectrometer.

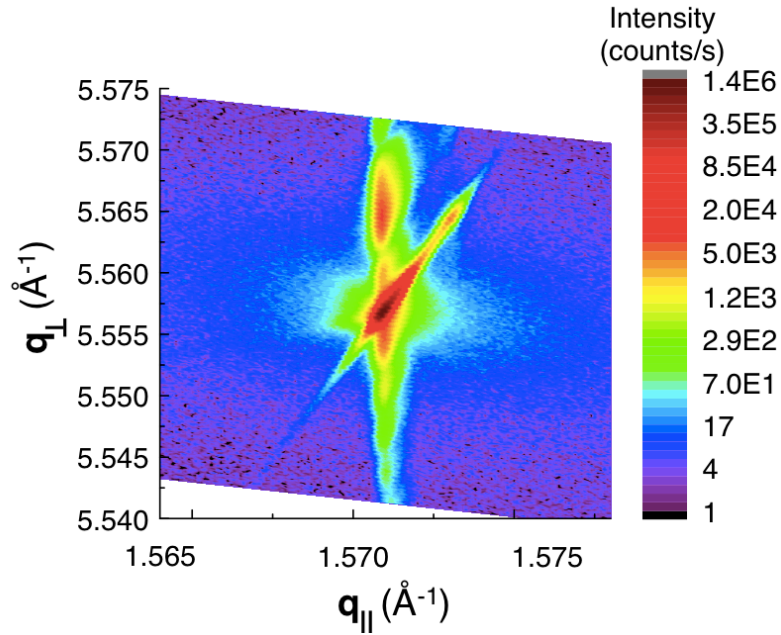

**Figure S2: Reciprocal space map of the GaAs and InGaAsN (115) peaks.**
Measurements were taken by aligning on the GaAs (115) peak, then varying 2θ and ω while keeping ϕ constant. The diagonal streak is an artifact from the monochromator, and the small, intense peak along this streak is the GaAs (115) Cu $K_{\alpha 2}$ peak. The intense peak above the GaAs (115) peak at $q_\perp = 5.564$ Å$^{-1}$ is the InGaAsN (115) peak, and the subsidiary maxima extending vertically along a line through this peak are thickness fringes from the InGaAsN layer[9, 10]. Since the InGaAsN (115) peak is directly above the GaAs (115) peak, the InGaAsN layer is pseudomorphic, which means the InGaAsN layer shares the same in-plane lattice parameter as GaAs.

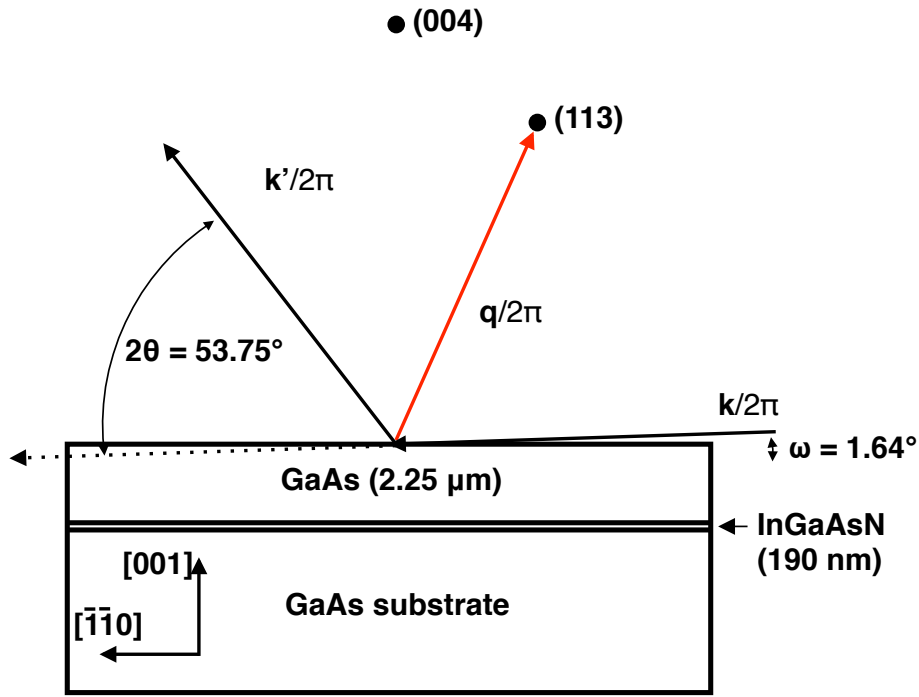

**Figure S3: Grazing incidence scattering geometry.** Schematic of the grazing incidence scattering geometry used for reciprocal space mapping the GaAs (113) peak. In this geometry, $\phi$ is fixed while $2\theta$ and $\omega$ are varied. The attenuation length of Cu $K_\alpha$ radiation in GaAs is 27 µm, calculated from values presented in Cullity[9, 10]. The incident angle to probe the GaAs (113) peak is 1.64°, and since the GaAs layer thickness is 2.25 µm, this corresponds to a path length of 78.6 µm until the InGaAsN layer is reached. Therefore, with this geometry, only 5.4% of the X-ray beam reaches the InGaAsN layer, and even less reaches the GaAs substrate. Therefore, the vast majority of diffracted intensity comes from the GaAs film, and a negligible amount of diffracted intensity comes from the InGaAsN layer and GaAs substrate.

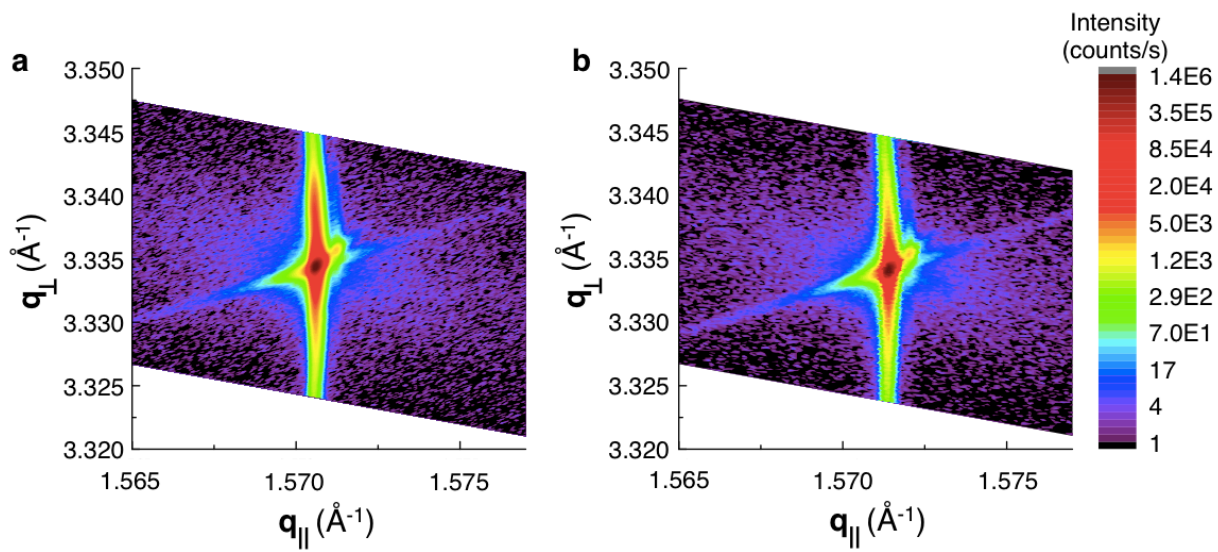

**Figure S4: Grazing incidence reciprocal space map of the GaAs film (113) peak.**
**a**, Grazing incidence reciprocal space map of the GaAs film (113) peak from a sample comprised of <001> GaAs substrate/InGaAsN (190 nm)/GaAs (2.25 µm). The narrow, intense peak is the GaAs film (113) peak, and the lower intensity peak above the GaAs film (113) peak is due to a small amount of diffracted intensity from the InGaAsN layer (InGaAsN (113) peak). **b**, Grazing incidence reciprocal space map of the GaAs (113) peak of a bare <001> GaAs substrate sawn from the same ingot as the growth substrate used in (**a**). The narrow, intense peak is the GaAs substrate (113) peak. In both reciprocal space maps, the diagonal streak running through the GaAs (113) peak is an artifact from the monochromator, and the lower intensity peak to the right of, and slightly above, the GaAs (113) peak is the GaAs (113) Cu $K_{\alpha 2}$ peak. The vertical streaks in each figure are surface effects caused by crystal truncation rods[13]. Both reciprocal space maps were taken by aligning the diffractometer on the GaAs (113) peak, and then varying 2θ and ω while keeping ϕ fixed.

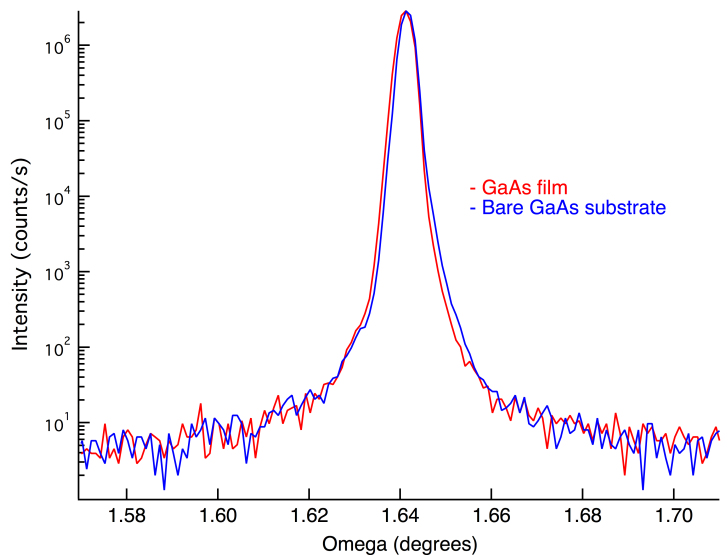

**Figure S5: GaAs film and substrate XRD rocking curves.** Grazing incidence rocking curves of the GaAs film (113) peak and of a bare GaAs substrate (113) peak. The GaAs film sample is comprised of <001> GaAs substrate/ InGaAsN (190 nm)/GaAs (2.25 µm), and the bare substrate sample is a <001>GaAs substrate sawn from the same ingot as the growth substrate used in the aforementioned heterostructure. Both rocking curves have a FWHM of 0.003° in ω, which indicates that the GaAs film is comparable in structural quality to the bare GaAs substrate[9, 10]. Both rocking curves were extracted from the reciprocal space maps shown in Figure S4 that were produced using the grazing incidence scattering geometry shown in Figure S3.

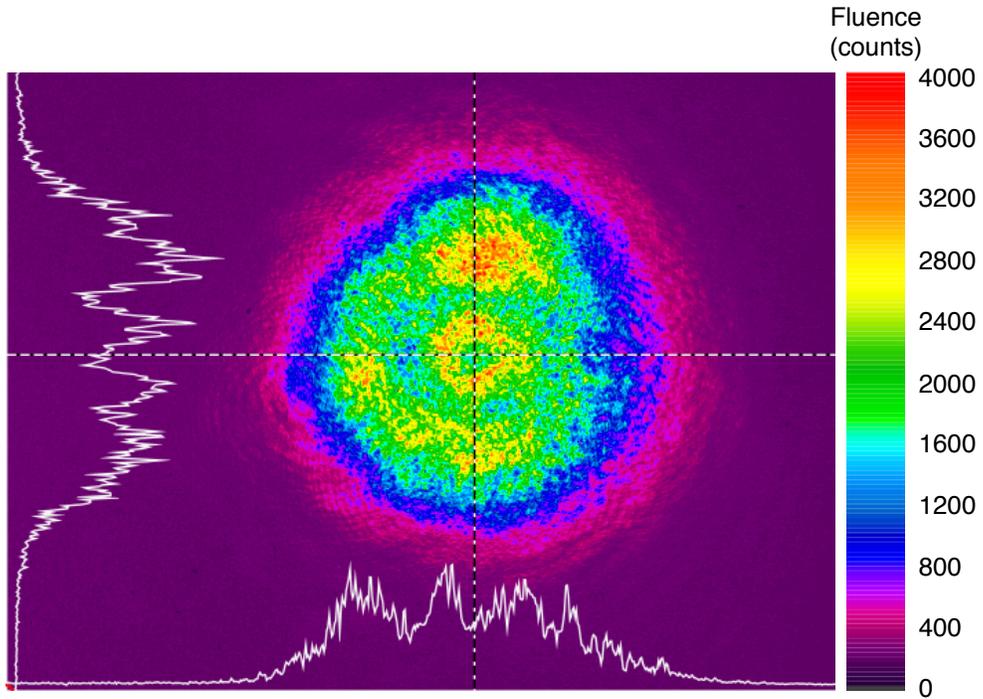

**Figure S6: Laser beam profile.** Fluence profile of the Q-switched Nd: YAG laser (Quanta-Ray model GCR-170) beam used for laser liftoff. The graphs on the horizontal and vertical axes show fluence (linear scale) at locations within the beam along the horizontal and vertical dotted lines. From this image, we see that the laser beam is highly inhomogeneous, with fluences varying by a factor of ~4 between the lowest fluence regions and the highest fluence regions.